# Defects in h-BN tunnel barrier for local electrostatic probing of two dimensional materials

Ying Liu[1,2], Zhenbing Tan[2], Manohar Kumar[2,3], T. S. Abhilash[2], Guan-jun Liu[1,a)] and Pertti Hakonen[2,b)]

[1]*Science and Technology on Integrated Logistics Support Laboratory, National University of Defense Technology, No. 109 Deya Road, 410073, Changsha, P. R. China*

[2]*Low Temperature Laboratory, Department of Applied Physics, Aalto University, Puumiehenkuja 2 B, 02150, Espoo, Finland*

[3]*Laboratoire Pierre Aigrain, Ecole Normale Superieure-PSL Research University, 24 rue Lhomond, 75231, Paris, France*

Defects in hexagonal boron nitride (h-BN) layer can facilitate tunneling current through thick h-BN tunneling barriers. We have investigated such current-mediating defects as local probes for materials in two dimensional heterostructure stacks. Besides *IV* characteristics and negative differential conductance, we have characterized the electrical properties of h-BN defects in vertical graphene-h-BN-Cr/Au tunnel junctions in terms of low frequency current noise. Our results indicate a charge sensitivity of $1.5 \times 10^{-5}$ e$/\sqrt{\text{Hz}}$ at 10 Hz, which is equal to good metallic single electron transistors. The noise spectra at low frequency are governed by a few two-level fluctuators. For variations in electrochemical potential, we achieve a sensitivity of 0.8 μeV$/\sqrt{\text{Hz}}$.

---

a) Electronic mail: gjliu342@126.com
b) Electronic mail: pertti.hakonen@aalto.fi

Van der Waals (VdW) heterostructures based on graphene and hexagonal boron nitride (h-BN) have attracted immense attention[1-3] in recent years both in fundamental physics[4-7] as well as for technological applications[8-9]. Hexagonal BN, structurally isomorphic to graphene with a ~2% lattice mismatch and with a large bandgap of ~5.9 eV, has intriguing potential for use as a dielectric layer in functional heterostructures[10, 11], in particular, as atomically smooth tunnel barriers[8,12]. Various spectroscopic studies have been employed to unravel electronic and phononic properties of h-BN, which are influenced by intrinsic nanometer-scale defect states in exfoliated h-BN crystals [13-15]. The defect states have been observed to enhance tunneling current across h-BN barriers in VdW devices, and clear Coulomb staircase patterns have been reported, indicating quantum dot behavior of the defect states[15].

A quantum dot, weakly coupled to source and drain contacts via tunnel barriers, forms a very sensitive electrometer due to Coulomb blockade effects. Such a single electron tunneling (SET) device remains as one of the most sensitive electrometers to date[16]. The defect states in an atomically thin h-BN barrier, may serve as a generic source of Coulomb islands, which provides a new way to study and probe local charge distribution and its dynamics in two dimensional materials.

In this work, we utilize CVD h-BN to construct vertical $Si^{++}$/$SiO_2$/graphene/h-BN/metal VdW heterostructures, in which defect states act as sensitive electrometers (for selection of the barrier material, see Supplementary Information). In our electrometer, graphene is used to replace the conventional metallic bottom electrode, owing to which the electrical field of the $Si^{++}$ bottom gate can penetrate through the intermediate graphene electrode and work on the defect states in the h-BN barrier. The observed sharp features due to Coulomb blockade and their gate modulation prove the existence of lattice defect states as quantum dots in the CVD h-BN. The performance of these quantum dots is characterized by measurements of low frequency current noise.

A schematic illustration of the vertical $Si^{++}$/$SiO_2$/graphene/h-BN/metal electrometer device is shown in Fig. 1a. The fork-shaped superconducting loop with graphene as a bridge filling the gap between the fork arms (350~400 nm apart) works as the source electrode. A few-layer CVD h-BN (Graphene Laboratories Inc., Lot# A041316) is inserted as tunnel barrier between the graphene and the top metal probe (drain electrode). Fig. 1b shows a false color Scanning Electron Micrograph (SEM) of the fabricated device. The green flake depicts graphene. The CVD h-BN is shown in blue, which covers almost the whole chip. According to earlier works[12-15], it is likely that at least one defect state is located in the h-BN right beneath the probe. The nature of the current flow from graphene to the probe (drain) is indicated using an orange dashed line. The current consists of two parts: 1) tunneling current through the h-BN barrier without the assistance of the defects (red arrow) and 2) sequential tunneling of electrons which first hop into a defect island and then tunnel into the probe (blue arrows). Owing to poor screening, defect states in the vertical graphene/h-BN/metal device can work as a SET electrometer even for charges trapped in SiO$_2$ below the graphene, which is indicated in Fig. 1c.

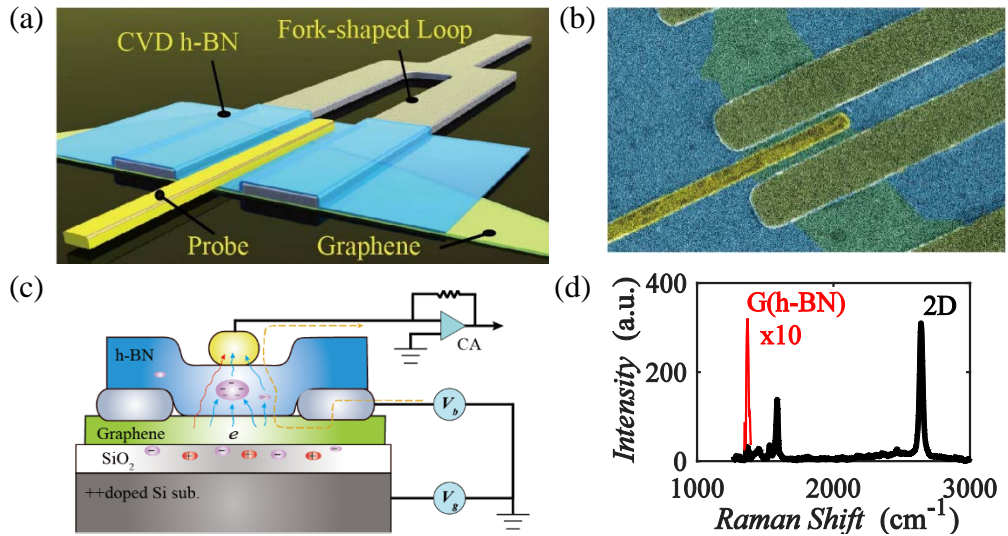

**Fig. 1.** a) Schematic of the vertical Graphene/h-BN/metal device: The blue is for h-BN, green for graphene, white for Al loop, and yellow for Au probe. b) False color SEM image of a fabricated device: Green color denotes the graphene flake. CVD h-BN is covering the whole sample and the probe is placed between the gap on top of the h-BN. c) Schematic view of transport via barrier defects in h-BN and the measurement configuration: The bias voltage ($V_b$) is applied to the loop and the current is read out from the current pre-amplifier (CA). The ++doped silicon substrate acts as the back gate at the electric potential $eV_g$. When the local potential changes, electrons hop from the graphene sheet to the probe electrode via barrier defects. The yellow dashed line denotes the path of flowing electrons. d) Typical Raman spectrum of a graphene/h-BN heterostructure: h-BN exhibits a characteristic G peak at ~1369 cm$^{-1}$ in the zoom-in red curve. Graphene exhibits a 2D peak without sideband at ~2640 cm$^{-1}$ as indicated in the figure.

The fabrication of our devices proceeded as follows. First, graphene was exfoliated and deposited on a heavily doped silicon substrate capped with thermally grown, 285 nm-thick $SiO_2$. Graphene stripes with one dimension in the range of 2~5 μm were preferred so that reactive ion etching (RIE) of graphene could be avoided, which also helped in reducing the processing residues on graphene surface. Next, a fork-shaped loop, consisting of 5 nm Ti and 45 nm Al, was patterned to contact graphene using standard electron-beam lithography and electron-beam evaporation techniques. Stacking of h-BN on graphene was based on delamination using polydimethylsiloxane (PDMS) stamp. After coating a 165 μm-thick PDMS film on the CVD h-BN/Copper, the copper was removed using 1 mol/L $FeCl_3$ solvent. To further remove inorganic and insoluble organic residuals, we used the modified RCA method by Liang *et al* to clean the h-BN[17, 18]. A 3-axes micromanipulator was employed to bring the stack close to graphene, which could be monitored by a change of color on the video screen. The manual approach using the manipulator was stopped sufficiently far to avoid breaking of the graphene sheet or the h-BN film by hand-induced vibrations. By taking advantage of the viscoelasticity of the PDMS, we controlled the movement of the stamp by adjusting the temperature of the heater on the transfer stage. By using such thermal-drive without vibrations, h-BN transfer fully overlapping the graphene could be achieved. This thermal-drive dry transfer method facilitates almost perfect large area transfer of h-BN without cracks. The method can be adapted for similar transfer of other 2D materials such as graphene, $MoS_2$ etc. Finally, a 200-nm-wide electrode consisting of 5 nm Cr and 45 nm Au was deposited right between the arms of the loop and on top of h-BN as a tunnel probe. The overlapped probe junction area in Fig. 1b is about $0.18 \times 1.5$ μm$^2$.

The Raman spectrum of h-BN on graphene is shown in Fig. 1d. Besides monolayer graphene 2D peak at 2640 cm$^{-1}$, the characteristic h-BN G-peak around 1369 cm$^{-1}$ due to the $E_{2g}$ phonon mode is clearly visible. For monolayer h-BN, the intensity of the G-peak is expected to be very weak[19], and it grows with increasing number of layers of h-BN. Hence, clear visibility of the G-peak indicates the h-BN in our device has several layers. This is consistent with our atomic force microscopy and tunneling experiments. The thickness of h-BN measured by atomic force microscopy is about 1.5 nm, suggesting 4 or 5 layers. Our tunneling resistance at room temperature is about 27 MΩμm$^2$ which is in the range 10 - 300 MΩμm$^2$ measured for 4 layers of h-BN[20, 21, 22]. Even though there are defects involving in the transport at room temperature, the intrinsic tunneling resistance at low temperature is less than 300 MΩμm$^2$. Therefore, we believe that the thickness of our h-BN corresponds to 4 layers.

Detailed electrical transport measurements were carried out at 20 mK using a Bluefors LD250 dilution refrigerator. The main focus was on low-bias measurements in which tunneling spectroscopy provides information on the role of impurities in tunneling transmission.

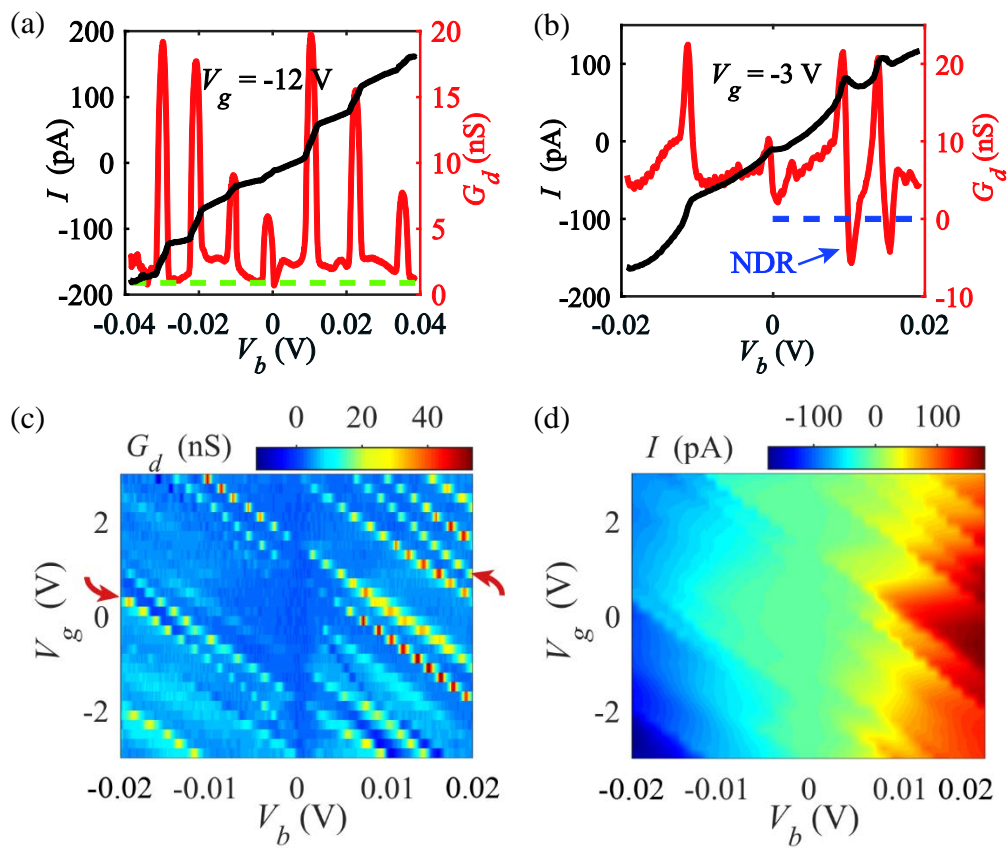

**Fig. 2.** a) and b) Staircase-like tunneling current as a function of bias voltage $V_b$ and the differential conductance $G_d = dI/dV_b$ vs. $V_b$ measured at $V_g = -12$ V and $V_g = -3$ V, respectively. The lock-in amplifier frequency was $f = 7$ Hz. The green dashed line indicates a nonzero $G_d$ background of about 1 nS. The $G_d$ below the blue dashed line becomes negative. c) Color map of differential conductance $G_d(7\text{Hz})$ plotted as a function of $V_g$ and $V_b$. Two of the borderlines with different slope from the others, marked by the red arrows, are assigned to a defect state, which indicates that there is more than one defect state in the h-BN layer. d) Color map of the tunneling current plotted as a function of $V_g$ and $V_b$. Both measurements were done at $T = 20$ mK.

The low-temperature current-voltage characteristics in Fig. 2a displays clear discrete steps, the size of which varies with the bias voltage $V_b$. In line with the steps of the $I$-$V$ curve, quasi-periodic oscillations in the differential conductance $G_d = dI/dV_b$ are shown. These observations are very similar to the Coulomb staircase features found in nanostructures[23]. The Coulomb staircase constitutes a general case where the tunnel barriers that isolate the island from the leads are asymmetric. Altogether, our data are similar to findings in exfoliated h-BN by the Eisenstein group[15, 22]. Spectrometric analysis[24-25] and atomic topography characterizations[14, 26-28] have convincingly demonstrated the prevalence of defects in h-BN flakes. These studies have revealed that, the principal defects are substitutional C at the N or B site ($C_N$ and $C_B$), O at the N site ($O_N$) and vacancy-type interstitials with passivated H (5|7 defect, 4|8 defect). In particular, C atoms are only seen to substitute for B–N pairs and they can easily form a full hexagonal C ring in h-BN crystal lattice. Considering the relatively large size of the defect island (3.5 nm) estimated in the following section from our data, we attribute the observed tunneling characteristics to metallic-like defect clusters in the h-BN insulator. The clusters may originate *e.g.* from substitutional carbon due to the growth process since the carbon impurities usually display a relatively large concentration compared to other type of defects. As shown by the green dashed line in Fig. 2a, $G_d$ has a nonzero background of about 1 nS, which is caused by the intrinsic tunneling through the h-BN without any help of defects. We notice that this intrinsic tunneling conductance varies with back gate voltage. This is likely because the density of states of the graphene is tuned by the back gate voltage, which affects the tunneling resistance.

As in Ref 22, we model the defect island as a disk of radius $r$ midway between the electrodes. Using the parallel plate approximation, the capacitance of such a disk can be estimated as $C_\Sigma = 3.2$ aF for a 3.5 nm radius disk in the middle of 1.5 nm thick h-BN[29]. We chose the value of the dielectric constant $\varepsilon = 3.5$, which corresponds to the average of the spread of values found in the literatures[30-32]. According to theory[33], the level spacing in such a dot is ~10 meV. The Coulomb energy $e^2/C_\Sigma$ accounts to 50 meV, which together with the level spacing suggests $50-60$ meV spacing of the resonances. On the $V_b$ axis, this is close to the large-scale periodicity seen in the pattern of $G_d$ in Fig. 2a. The estimated dimension of the defect island in this work is pretty well consistent with the STM topography measured in Ref 14.

Fig. 2c displays $G_d$ as a function of the bias $V_b$ and the back gate voltage $V_g$. The Coulomb diamonds are not visible in Fig. 2c even though the staircase borderlines are very prominent (The grainiess of the negative-slope lines is due to large gate spacing $\Delta V_g = 0.2$ V in the measurement). Most of the bright lines with a maximum in $G_d$ are parallel to each other except for two, which indicates that there is more than one defect state in the h-BN layer. We assign these two distinct lines, separated in gate voltage by $\Delta V_g = 6.4$ V, to a single defect. Assuming this value for $\Delta V_g$ as representative of all defects, we obtain $C_g = 2.5 \times 10^{-20}$ F for the average capacitance between the defect and the back gate electrode. Thus, we obtain for the gate lever arm $\alpha = C_g/C_\Sigma = 8 \times 10^{-3}$, which yields for the approximate energy level spacing $(C_g/C_\Sigma)\, e\Delta V_g \simeq 50$ meV. By counting the resonances in between the lines with distinct slopes, we find $N = 7$ for the number of defects active in the sample. The irregular periodicity of the conductance oscillations as a function of $V_g$ is a consequence of several, nearly equivalent defects. Close to $V_b = 0$, the conductance is suppressed, and a narrow energy gap appears across the defect resonances $G_d(V_b, V_g)$. We speculate that this might be due to Coulomb blockade effects in the graphene flake, in which single carriers would be localized by the superconducting aluminum leads. Clearly, the gap is too large for a proximity-induced mini-gap[12, 34, 35].

A full staircase map $I(V_b, V_g)$ is depicted in Fig. 2d. In accordance with the data in Fig. 2c, no Coulomb diamonds can be identified in Fig. 2d. This kind of behavior is common to SET devices in which the junction conductance differs by an order of magnitude or more. The transport cusps in the SET current arise only due to the change in transport channels available in the better conducting junction, and the cusps due to the other junction will remain small. Furthermore, we note that, in the presence of seven nearly equivalent impurity islands, the size of the Coulomb diamond for one single island would exceed the scale of the employed bias voltage.

Below the blue dashed line in Fig. 2b, the differential conductance $G_d$ becomes negative while in the *I-V* curve, the corresponding $I$ decreases with $V$. Negative differential conductance, caused by in-plane band alignment and phonon scattering, has been widely reported in h-BN tunneling barriers[36-39]. In Ref 40, the phonon-assisted electron tunneling peaks in the conductance (at energies ~12 - 200 meV) were found to be independent of the gate voltage. However, the differential conductance peaks observed in our experiments are gate-dependent and shift continuously with the bias voltage. Hence, the regular phonon-assisted processes can be ruled out. In our experiment, the negative differential conductance appearing with $V_b$ and $V_g$ in between the $G_d$ peaks in Fig. 2c implies that it is related with the defects. Therefore, we believe the negative differential conductance in our case is caused by interaction between defects, which is different from the previous reports on h-BN[36-39].

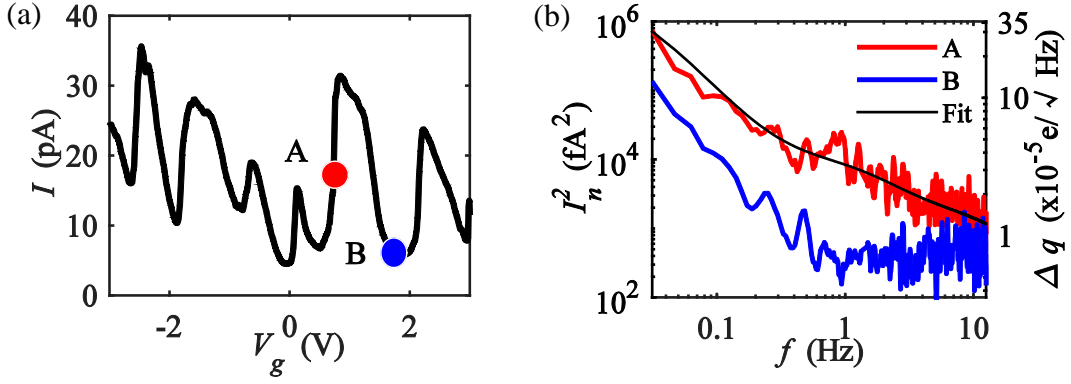

**Fig. 3** a) Gate modulation of current at $V_b = 5$ mV. Points marked by A and B denote the biasing conditions for the noise measurements at maximum and zero gain, respectively. b) Current noise $I_n^2$ measured at minimum (lower) and maximum gain (upper trace). The right axis is obtained by using the inverse of charge conversion gain $G_{Iq} = \Delta I / \Delta q$ for scaling, which gives the equivalent charge noise for the upper trace. The black curve shows the fitting result using a sum of three Lorentzian spectra corresponding to three independent two-level fluctuators. The calibration of the charge noise scale was obtained using a 7 Hz AC-signal of 4 mV amplitude applied directly to the gate.

Sharp current steps along the $V_g$ direction in Fig. 2d indicate exceptional potential for sensitive electrometry. For more accurate characterization, current modulation $I(V_g)$ is displayed in Fig. 3a at $V_b = 5$ mV. The slope of the curve $k = \frac{dI}{dV_g} = \frac{1}{c_g}\frac{\delta I}{\delta q}$ provides conversion gain $G_{Iq} = \delta I / \delta q$ for charge ($q$) detection, while the rounded bottom forms a charge insensitive regime. At the point marked by A, we obtain the maximum slope of $k \approx 0.44$ nA/V. The insensitive point marked by B is used as a reference. The conversion gain for charge $G_{Iq} \approx k\, \Delta V_g/e = 2.8$ nA/$e$ is obtained at point A.

The current noise measured at both the maximum and minimum conversion gain of the electrometer (points A and B) is displayed in Fig. 3b. The frequency dependence of the current noise is close to $1/f^\alpha$ with $\alpha \sim 2$, as observed in many nanodevices with a single two-level fluctuator[41]. Our charge noise data at point A can be fit using a sum of three Lorentzian spectra: $S_I = \sum_{i=1}^{3} \frac{S_{Li}\tau_i}{1+\omega^2\tau_i^2}$, where $1/\tau_i$ denotes the sum of up and down transition rates of fluctuator $i$, and $S_{Li}$ specifies its spectral weight. An excellent fit with a $R^2$-value close to 1 was obtained (see Supplementary Information for details). This means that our device contains two-level systems, charge traps, presumably at the $SiO_2$/graphene interface, which modulate the tunneling rates of the defect quantum dots in h-BN. At a frequency of $f = 10$ Hz, we obtain a good charge sensitivity of $1.5 \times 10^{-5}$ e/$\sqrt{Hz}$, which is comparable with the record performance $\sim 6 \times 10^{-6}$ e/$\sqrt{Hz}$ at 45 Hz in carbon nanotube SETs[42]. The fluctuators also lead to substantially enhanced noise from the theoretically estimated minimum level of $\delta q_{min} \sim 1 \times 10^{-6}$e/$\sqrt{Hz}$[43]. Even though the noise level of our $Si^{++}$/$SiO_2$/graphene/h-BN/metal "single defect" SET is one order of magnitude higher than the theoretical limit, the performance is already much better than for conventional FET-based electrometers. Using the gate capacitance $C_g$ and the lever arm ratio $C_g/C_\Sigma = 0.008$ to convert the charge sensitivity into a resolving power for electrical potential, we obtain a sensitivity of 0.8 $\mu$eV/$\sqrt{Hz}$ for electrochemical potential. This is slightly better than achieved using scanning SET probe[44, 45].

In summary, we have characterized a defect-based SET within a graphene/h-BN/metal heterostructure, which was fabricated using a dry transfer method with thermal position control for transferring a CVD h-BN tunneling barrier onto graphene. We observe staircase-like features in *I-V* characteristics, and the gate modulation of the conductance peaks indicates single electron tunneling events via intrinsic defect states in CVD h-BN. Our device works as a SET-based electrometer, in which only one defect at a time is active at low bias. The current noise measurements demonstrate a good charge sensitivity, as well as sensitivity to local electrostatic potential. This type of SET opens up new possibilities for ultra-sensitive electrometry for various applications, in particular for local studies of 2D materials, both in heterostructures and quantum Hall states at large magnetic field.

## Acknowledgment

We acknowledge fruitful discussions with R. N. Jabdaraghi and J. Sarkar. Our work was supported by the Academy of Finland (Contracts 310086 and 314448) and by the European Union under Grant Agreement No. 696656 Graphene Flagship. Our work was also supported by the